\documentclass[12pt]{article}
\usepackage{epsfig}
\usepackage{hyperref}%

\date{}
\begin{document}

\newcommand{\beq}{\begin{equation}}
\newcommand{\eeq}{\end{equation}}
\newcommand{\nn}{\nonumber}
\newcommand{\bea}{\begin{eqnarray}}
\newcommand{\eea}{\end{eqnarray}}

\title{Brane Dynamics, the Polytropic Gas and Conformal Bulk Fields}

\author{Rui Neves\footnote{E-mail: \tt
rneves@ualg.pt}\hspace{0.2cm}
  and
Cenalo Vaz\footnote{E-mail: \tt cvaz@ualg.pt}\\
{\small \em Faculdade de Ci\^encias e Tecnologia,
Universidade do Algarve}\\
{\small \em Campus de Gambelas, 8000-117 Faro, Portugal}
}

\maketitle

\begin{abstract}
We consider in the Randall-Sundrum scenario the dynamics of
a spherically symmetric 3-brane world when matter fields exist in the
bulk. We determine exact 5-dimensional solutions which
localize gravity near the brane and
describe the dynamics of homogeneous polytropic matter on the
brane. We show that these geometries are associated with a well
defined conformal class of
bulk matter fields. We analyze the effective polytropic dynamics
on the brane identifying conditions which define it as singular or
as globally regular.
\vspace{0.5cm}


\end{abstract}

\section{Introduction}

In the Randall-Sundrum (RS) scenario \cite{RS1,RS2} the Universe we inhabit
is a 3-brane world embedded in a $Z_2$ symmetric 5-dimensional
anti-de Sitter (AdS) space. With two branes the
scenario provides a solution to the hierarchy problem introducing an
exponential warp in the fifth dimension \cite{RS1}. Gravity
is localized on the hidden positive tension brane and decays
towards the visible brane which has negative tension. This leads to an
exponential hierarchy between the weak and Planck scales and also
to TeV scale Kaluza-Klein mass splittings and
couplings \cite{RS1}. A bulk scalar field may be
used to stabilize the size of the fifth dimension \cite{GW,WFGK} and generate
the Friedmann-Robertson-Walker (FRW) cosmology on the visible brane
\cite{CGRT}. In addition it allows the low energy recovery
of 4-dimensional Einstein gravity on the visible brane \cite{TM}. Two
branes with positive tensions and an infinite fifth dimension may also
be considered \cite{LR}. The exponential hierarchy is then obtained on
the brane where the gravitational field is not localized and it turns
out that for a low energy observer gravity is effectively described by
4-dimensional general relativity \cite{LR}. A warped infinite fifth
dimension is also associated with a single positive tension brane
to which the gravitational field is bound \cite{RS2}. Again the low
energy theory of gravity on the brane is
effectively 4-dimensional general relativity \cite{RS2,GT,GKR} and the
brane cosmology may be of the FRW type \cite{NK} (see also \cite{KKOP} where
a single brane in a compact fifth dimension was discussed). The
gravitational collapse of matter localized on a 3-brane was also
addressed in this model \cite{CHR}-\cite{RC}. To date
realistic 5-dimensional collapse solutions remain to be found. The difficulty
resides in achieving a simultaneous localization of
gravity and matter which avoids the creation of naked singularities
in the bulk. In this context it should be noted that exact
solutions which may be interpreted as static black holes
localized on a brane have been found for a 2-brane embedded in a
4-dimensional AdS bulk \cite{EHM}. On the contrary, the exact
5-dimensional solutions discovered so far display singularities
somewhere in the bulk \cite{CHR,KOP}, \cite{LP}-\cite{RC}.

In this letter we proceed with the investigation about the dynamics of a
spherically symmetric RS 3-brane when conformal matter fields exist in
the bulk \cite{RC}. We analyze the possibility of generating on the
brane the dynamics of a perfect fluid with a polytropic equation of
state. We dedicate special attention to its generalized Chaplygin
phase which recently has been considered to
be a viable alternative to model the accelerated expansion of the
Universe \cite{KMP}-\cite{BBS}.

\section{Einstein Equations and Conformal Bulk Matter}

Let us map the RS orbifold by a set of
comoving coordinates $(t,r,\theta,\phi,z)$. The non-factorizable
dynamical metric consistent with the $Z_2$ symmetry in $z$ and with
4-dimensional spherical symmetry in $(t,r,\theta,\phi)$ may be written
as

\beq
d{\tilde{s}_5^2}={\Omega^2}d{s_5^2},\quad d{s_5^2}=-{e^{2A}}d{t^2}
+{e^{2B}}d{r^2}+{R^2}d{\Omega_2^2}+d{z^2}.
\eeq
The functions $\Omega=\Omega(t,r,z)$, $A=A(t,r,z)$, $B=B(t,r,z)$ and
$R=R(t,r,z)$ are $Z_2$ symmetric. $\Omega$ is the RS warp factor and
$R$ represents the physical radius of the 2-spheres. The classical dynamics
is defined by

\beq
{\tilde{G}_\mu^\nu}=-{\kappa_5^2}\left[{\Lambda_B}{\delta_\mu^\nu}+
{\lambda\over{\sqrt{\tilde{g}_{55}}}}\delta
\left(z-{z_0}\right)\left(
{\delta_\mu^\nu}-{\delta_5^\nu}{\delta_\mu^5}\right)-
{\tilde{T}_\mu^\nu}\right],
\label{5DEeq}
\eeq
where $\Lambda_B$ is the negative bulk cosmological constant,
$\lambda$ is the brane tension and ${\kappa_5^2}=8\pi/{M_5^3}$ with $M_5$ the
fundamental 5-dimensional Planck mass. The brane is located at
$z={z_0}$  and the stress-energy tensor ${\tilde{T}_\mu^\nu}$ is
conserved in the bulk, ${\tilde{\nabla}_\mu}{\tilde{T}_\nu^\mu}=0$.

Let us consider the special class of conformal bulk matter defined by
${\tilde{T}_\mu^\nu}={\Omega^{-2}}{T_\mu^\nu}$ and assume that $T_\mu^\nu$
depends only $(t,r)$. The matter represented by ${\tilde{T}_\mu^\nu}$ is
then not localized near the brane. Instead, because the conformal
weight is $-4$, the density and pressures increase with $z$ and eventually
become singular at the AdS horizon. The same problem is found in
solutions with an extended black string singularity \cite{CHR},
\cite{LP,MPS,KT}. The divergence of ${\tilde{T}_\mu^\nu}$ makes it necessary to
introduce two branes with opposite tensions in the bulk, which will intersect the
space-time before the AdS horizon is reached. This was suggested in the
first RS model \cite{RS1} and the two branes will then have identical cosmological
evolutions, though gravity will be localized on the first brane but not on the
second. In a single brane model one must look for solutions requiring
a simultaneous localization of matter and gravity in the vicinity of
the brane \cite{KT}. Assuming further that $A=A(t,r)$, $B=B(t,r)$,
$R=R(t,r)$ and $\Omega=\Omega(z)$ we obtain \cite{RC}
\beq
{G_a^b}={\kappa_5^2}{T_a^b},\;{\nabla_a}{T_b^a}=0,\;
{G_z^z}={\kappa_5^2}{T_z^z},\label{4DEeq}
\eeq
and
\[
6{\Omega^{-2}}{{({\partial_z}\Omega)}^2}=
-{\kappa_5^2}{\Omega^2}{\Lambda_B},
\]
\beq
3{\Omega^{-1}}{\partial_z^2}\Omega=-{\kappa_5^2}{\Omega^2}
\left[{\Lambda_B}+\lambda{\Omega^{-1}}\delta(z-{z_0})\right].\label{rswf}
\eeq
where the latin index represents the coordinates $t,r,\theta$ and
$\phi$. If the stress-energy tensor is
${T_\mu^\nu}=diag(-\rho,{p_r},{p_T},{p_T},{p_z})$, where
$\rho$, $p_r$, $p_T$ and $p_z$ denote the bulk matter density and
pressures, the non-trivial Einstein equations in Eq. (\ref{4DEeq}) are given by

\beq
{G_t^t}={e^{-2B}}\left({{{R'}^2}\over{R^2}}+2{{R''}\over{R}}
-2B'{{R'}\over{R}}\right)-{1\over{R^2}}
-{e^{-2A}}\left({{\dot{R}^2}\over{R^2}}+2\dot{B}{{\dot{R}}\over{R}}\right)
=-{\kappa_5^2}\rho,\label{4DEeq1}
\eeq
\beq
{G_t^r}={{2{e^{-2A}}}\over{R}}\left(\dot{R}'-A'\dot{R}-\dot{B}R'\right)=0,
\label{4DEeq2}
\eeq
\beq
{G_r^r}={e^{-2B}}\left({{{R'}^2}\over{R^2}}+2A'{{R'}\over{R}}\right)
-{1\over{R^2}}-{e^{-2A}}\left({{\dot{R}^2}\over{R^2}}+2{{\ddot{R}}\over{R}}
-2\dot{A}{{\dot{R}}\over{R}}\right)={\kappa_5^2}{p_r},\label{4DEeq3}
\eeq
\bea
{G_\theta^\theta}={G_\phi^\phi}&=&{e^{-2B}}\left({{A'}^2}+A''-A'B'
+A'{{R'}\over{R}}+{{R''}\over{R}}-B'{{R'}\over{R}}\right)\nonumber\\
&-&{e^{-2A}}\left({\dot{B}^2}+\ddot{B}-\dot{A}\dot{B}+{{\ddot{R}}\over{R}}
-\dot{A}{{\dot{R}}\over{R}}+\dot{B}{{\dot{R}}\over{R}}\right)
={\kappa_5^2}{p_T},\label{4DEeq4}
\eea
and
\bea
{G_z^z}&=&{e^{-2B}}\left({{A'}^2}+A''-A'B'+{{{R'}^2}\over{R^2}}
+2A'{{R'}\over{R}}+2{{R''}\over{R}}-2B'{{R'}\over{R}}\right)-{1\over{R^2}}
\nonumber\\
&-&{e^{-2A}}\left({\dot{B}^2}+\ddot{B}-\dot{A}\dot{B}+2{{\ddot{R}}\over{R}}
+{{\dot{R}^2}\over{R^2}}-2\dot{A}{{\dot{R}}\over{R}}
+2\dot{B}{{\dot{R}}\over{R}}\right) = {\kappa_5^2}{p_z}.\label{5DEeqz}
\eea
where the dot and the prime denote, respectively, partial
differentiation with respect to $t$ and $r$. On the other hand the
conservation equations are
\beq
\dot{\rho}+\dot{B}\left(\rho+{p_r}\right)+2{{\dot{R}}\over{R}}
\left(\rho+{p_T}\right)=0,\; A'\left(\rho+{p_r}\right)+{p_r}'
-2{{R'}\over{R}}\left({p_T}-{p_r}\right)=0\label{4Dceq0}
\eeq
and we also find \cite{RC} that $p_z$ must satisfy a trace
equation of state \cite{KKOP,IR}
\beq
\rho-{p_r}-2{p_T}+2{p_z}=0,
\eeq
for consistency.

For definiteness $\Omega$ may be taken to be
$\Omega={\Omega_{\mbox{\tiny RS}}}=l/(|z-{z_0}|+{z_0})$
where
$l=\sqrt{-6/({\Lambda_B}{\kappa_5^2})}$ and ${\Lambda_B}=-{\kappa_5^2}
{\lambda^2}/6$ \cite{RS1,RS2,CHR}. This warp factor also holds in the
two brane model as may be seen taking the periodicity and the $Z_2$
symmetry of the orbifold into account. Then Eq. (\ref{4DEeq}) must be
satisfied on both branes, in this sense twin universes with an identical
cosmological evolution.

\section{Polytropic Dynamics on the Brane}

To behave as a polytropic fluid interacting with an effective cosmological
constant on the brane, the diagonal conformal bulk matter should be
defined by

\beq
\rho={\rho_{\mbox{\tiny P}}}+\Lambda,\;
{p_r}+\eta{{\rho_{\mbox{\tiny P}}}^\alpha}+
\Lambda=0,\;{p_T}={p_r},\;{p_z}=-{1\over{2}}
\left({\rho_{\mbox{\tiny P}}}
+3\eta{{\rho_{\mbox{\tiny P}}}^\alpha}\right)
-2\Lambda,
\label{dmeqst}
\eeq
where ${\rho_{\mbox{\tiny P}}}$ defines the polytropic energy
density. $\Lambda$ is the bulk quantity which mimics the brane
cosmological constant. In the two brane model the
usual 4-dimensional quantities are given by
${\rho_4}={\rho_{\mbox{\tiny P}}}L$, ${\Lambda_4}=\Lambda L$ where
$L={\kappa_5^2}/{\kappa_4^2}$
is the length scale defining the interbrane distance and thus the size
of the extra dimension. The parameters ($\alpha$, $\eta$) characterize
different polytropic phases.

Let us begin by solving the conservation equations in Eq. (\ref{4Dceq0}).
Taking into account Eq. (\ref{dmeqst}) we write them as follows
\beq
\dot{{\rho_{\mbox{\tiny P}}}}+\left(\dot{B}+2{{\dot{R}}\over{R}}\right)
\left({\rho_{\mbox{\tiny P}}}-\eta{{\rho_{\mbox{\tiny P}}}^\alpha}\right)=0,\;
A'\left({\rho_{\mbox{\tiny P}}}-\eta{{\rho_{\mbox{\tiny P}}}^\alpha}\right)
-\eta\alpha{{\rho_{\mbox{\tiny P}}}^{\alpha-1}}{\rho_{\mbox{\tiny P}}}'=0.
\label{4Dceq}
\eeq
The contribution of the cosmological constant cancels out in
Eqs. (\ref{4Dceq}). Specializing to the case
of a homogeneous energy density, $\rho_{\mbox{\tiny P}} =
\rho_{\mbox{\tiny P}}(t)$, the metric function $A(t,r)$
can be safely set to zero. Combining with the off-diagonal Einstein
equation (\ref{4DEeq2}), which has solution ${e^B}=R'/H$ with $H=H(r)$
an arbitrary function of $r$, and separating the variables $t$ and $r$
we obtain \cite{BBS}

\beq
{\rho_{\mbox{\tiny P}}}={{\left(\eta+{a\over{S^{3-3\alpha}}}\right)}^
{1\over{1-\alpha}}},\label{dmdena}
\eeq
where $\alpha\not=1$, $a$ is an integration constant and $S=S(t)$ is the
Robertson-Walker scale factor of the brane world which is related to
the physical radius by $R=rS$. For $\alpha=1$
Eq. (\ref{dmdena}) is not defined. The appropriate limit is
${\rho_{\mbox{\tiny P}}}=b/{S^{3-3\eta}}$,
where $b$ is the corresponding integration constant and
$\eta\not=1$.

Next let us consider the diagonal Einstein equations in
Eq. (\ref{4DEeq}). First note that since ${p_r}={p_T}$ we must have ${G_r^r}=
{G_\theta^\theta}$, which is possible only if,
\beq
{H^2}=1-k{r^2},\label{HRWeq}
\eeq
where the constant $k$ is the Robertson-Walker curvature parameter. Using
Eqs. (\ref{4DEeq1}), (\ref{4DEeq3}), (\ref{4DEeq4}) and (\ref{dmeqst}) we
form the sum,
\beq
{-G_t^t}+{G_r^r}+2{G_\theta^\theta}=-2{{\ddot{R}'}\over{R'}}-
4{{\ddot{R}}\over{R}}={\kappa_5^2}\left({\rho_{\mbox{\tiny P}}}-3\eta
{{\rho_{\mbox{\tiny P}}}^\alpha}-2\Lambda\right).\label{4DEteq}
\eeq
Then substituting $R=rS$ we obtain
\beq
{\ddot{S}\over{S}}=-{{\kappa_5^2}\over{6}}\left({\rho_{\mbox{\tiny P}}}-3\eta{{\rho_{\mbox{\tiny
          P}}}^\alpha}-2\Lambda\right).\label{dmdyeq}
\eeq
Applying the radial equation (\ref{4DEeq3}) then leads to
\beq
{\dot{S}^2}={{\kappa_5^2}\over{3}}\left({\rho_{\mbox{\tiny
        P}}}+\Lambda\right){S^2}-k.\label{dmdyeqa}
\eeq
Now, Eqs. (\ref{dmdyeq}) and (\ref{dmdyeqa}) are linked by a
derivative. They are consistent with each other when ${\rho_{\mbox{\tiny P}}}$
obeys the conservation Eq. (\ref{4Dceq}).

Finally, introducing
Eqs. (\ref{dmdyeq}), (\ref{dmdyeqa}) and (\ref{HRWeq}) on
Eq. (\ref{5DEeqz}) and using the expression for $p_z$ given in
Eq. (\ref{dmeqst}) the $zz$ component of
the 5-dimensional Einstein equations given in Eq. (\ref{5DEeqz}) is
seen to be an identity for all the parameters of the model. We have
thus obtained the following 5-dimensional polytropic solutions
for which gravity is confined near the brane

\beq
d{\tilde{s}_5^2}={\Omega_{\mbox{\tiny RS}}^2}\left[-d{t^2}+{S^2}
\left({{d{r^2}}\over{1-k{r^2}}}+{r^2}d{\Omega_2^2}\right)+d{z^2}\right],
\label{dmsol1}
\eeq
where the brane scale factor $S$ satisfies
Eq. (\ref{dmdyeqa}). It is easy to verify that if a 4-dimensional
observer confined to the
brane makes the same assumptions about the bulk degrees
of freedom then she deduces exactly the same dynamics
\cite{RC}. In fact, the non-zero components of the projected Weyl
tensor \cite{SMS} are given by

\beq
{\mathcal{E}_t^t}={{\kappa_5^2}\over{4}}\left({\rho_{\mbox{\tiny
        P}}}-\eta{\rho_{\mbox{\tiny P}}^\alpha}\right)
,\quad
{\mathcal{E}_r^r}={\mathcal{E}_\theta^\theta}={\mathcal{E}_\phi^\phi}=
-{{\mathcal{E}_t^t}\over{3}}.
\eeq
As a consequence, the effective 4-dimensional dynamics is indeed defined by
${G_t^t}=-{\kappa_5^2}({\rho_{\mbox{\tiny P}}}+\Lambda),\;
{G_r^r}={G_\theta^\theta}={G_\phi^\phi}=-{\kappa_5^2}(\eta
{\rho_{\mbox{\tiny P}}^\alpha}+\Lambda)$. The 4-dimensional observer
also sees gravity confined to the brane since she measures a negative
tidal acceleration \cite{RM} given by ${a_T}={\kappa_5^2}{\Lambda_B}/6$.
Because of this the geodesics just outside the brane converge towards
the brane and so for the 4-dimensional observer the conformal bulk
matter is effectively trapped inside the brane.

At small $S$ and for $\alpha<1$ the polytropic dynamics is
dominated by the homogeneous Oppenheimer-Snyder phase with
${\rho_{\mbox{\tiny P}}}={a^{1/(1-\alpha)}}/{S^3}$ which corresponds to
$\eta=0$. If $\alpha\geq 1$ this is no longer true. For $\alpha=1$
the dynamics is only that of dust
if $\eta=2/3$. For $\eta=-1/3$ we find radiation. If $\alpha>1$ the
small $S$ dynamics is that of an effective cosmological constant. Note
also that for $-1\leq\alpha<0$
and $0<\alpha\leq 1$ there is an intermediate phase defined by the equation of
state ${p_{\mbox{\tiny P}}}=-\alpha{\rho_{\mbox{\tiny P}}}$ which
satisfies the dominant energy condition. If $|\alpha|>1$ this condition
is violated. For large $S$ the
dynamics is dominated by an effective cosmological constant term and
it corresponds to $a=0$ if $\alpha\not=1$ or $b=0$ if $\alpha=1$.

In what follows we consider $-1\leq\alpha<0$. This is the generalized Chaplygin
phase of the polytropic gas. To
analyze it \cite{RC1,DJCJ} let us re-write Eq. (\ref{dmdyeqa}) (with
Eq. (\ref{dmdena}) substituted in) to define the potential $V(S)$ as follows

\beq
S{\dot{S}^2}=V(S)={{\kappa_5^2}\over{3}}\left[{{\left(\eta{S^{3-3\alpha}}+
a\right)}^{1\over{1-\alpha}}}+\Lambda{S^3}\right]-k S.\label{dmdineq}
\eeq
Naturally, the dynamics depends on $\Lambda$, $k$, $\eta$, $a$ and
$\alpha$. Let us consider some special ilustrative examples which can
be treated analytically.

Starting from an initial state, a rebounce will occur whenever ${\dot S} = 0$
and $S\neq 0$. This happens when $V(S)=0$. To begin with consider $k=0$. Then
introducing $Y=S^{3-3\alpha}$ the potential $V$ is given by

\beq
V=V(Y)={{\kappa_5^2}\over{3}}\left[{{\left(\eta Y+
a\right)}^{1\over{1-\alpha}}}+\Lambda{Y^{1\over{1-\alpha}}}\right].
\label{dmdineq0}
\eeq
Next take $\eta>0$ and $a>0$ \cite{BBS}. If
$\Lambda>0$ then $V>0$ for all
$-1\leq\alpha<0$ and $S\geq 0$. Consequently, the Chaplygin
shells may either expand continously to infinity or collapse to the
singular epoch at $S=0$ where
$V(0)={\kappa_5^2}{{a}^{1/(1-\alpha)}}/3>0$. Note that this is also what
happens if $\Lambda=0$. If $\Lambda<0$ then two new possibilities
arise. For $\eta\geq
{{(|\Lambda|)}^{1-\alpha}}$ the Chaplygin gas is dominant
and the potential is positive for all $Y\geq 0$. This implies an
evolution of the type found for $\Lambda\geq 0$. If $\eta<
{{(|\Lambda|)}^{1-\alpha}}$ then it is the cosmological
constant which prevails. The allowed dynamical phase space is given by
$0\leq Y\leq{Y_*}$ where
${Y_*}=a/[{{(|\Lambda|)}^{1-\alpha}}-\eta]$ is the
maximum regular rebounce epoch. Since at $S=0$ the shells meet
the singularity these solutions are not globally regular on the brane.

It is important to note that the range $\eta<0$ and $a<0$ may be
allowed in the presence of a non-zero cosmological constant. The same
is not true for $\eta>0$ and $a<0$ or $\eta<0$ and $a>0$. Indeed, for
these definitions the current experimental bounds are violated
\cite{BBS} and there is a singular epoch at
$Y=-a/\eta>0$ where $V'$ diverges. If $\eta<0$ and $a<0$ the potential depends
on ${(-1)}^{1/(1-\alpha)}$ and it is not defined for all values of
$-1\leq\alpha<0$.

For $\alpha=-p/q$, $q>p$ with $q$ and $p$, respectively, even and odd
integers, the Chaplygin contribution to
the potential $V$ is positive for all $S\geq 0$ and so the evolution
follows the case $\eta>0$ and $a>0$. However, if $q$ is odd and $p$ is
even then the potential becomes

\beq
V=-{{\kappa_5^2}\over{3}}\left[{{\left(|\eta| Y+
|a|\right)}^{q\over{p+q}}}-\Lambda{Y^{q\over{p+q}}}\right]
\eeq
and new dynamics develops. For a rebounce, $\Lambda$ must be positive and
then the dynamical phase space is defined by
$Y\geq|a|/({{(\Lambda)}^{(p+q)/q}}-|\eta|)$,
${{(\Lambda)}^{(p+q)/q}}>|\eta|$,
where the lower limit of the interval is the only existing regular rebounce
point. In this phase the Chaplygin energy density is negative but the
presence of the cosmological constant ensures that the total
density $\rho$ is not. Since
$V(0)=-{\kappa_5^2}{{|a|}^{q/(p+q)}}/3<0$ the singularity
at $S=0$ is inside the forbidden region and does not form. These
solutions are thus globally regular on the brane.

Another special case is that of $\Lambda=0$. As we have
seen for $k=0$ there are only singular solutions without rebouncing
epochs. It is for $k\not=0$ that new dynamics appears. With $Y={Z^3}$
the potential $V$ is given by

\beq
V=V(Z)={{\kappa_5^2}\over{3}}{{\left(\eta {Z^3}+
a\right)}^{1\over{1-\alpha}}}-k{Z^{1\over{1-\alpha}}}.
\eeq
Consider $k>0$, $\eta>0$ and $a>0$. The condition $V\geq 0$
is equivalent to ${\mathcal{V}}={\mathcal{V}}(Z)\geq 0$ where

\beq
{\mathcal{V}}={{\left({{\kappa_5^2}\over{3}}\right)}^{1-\alpha}}
\left(\eta{Z^3}+a\right)-{k^{1-\alpha}}Z.
\eeq
Then there are no more than two regular rebounce epochs in the allowed
dynamical phase space. Since
${\mathcal{V}}(0)={{({\kappa_5^2}/3)}^{1-\alpha}}a>0$
and ${\mathcal{V}}'=3{{({\kappa_5^2}/3)}^{1-\alpha}}\eta{Z^2}-{k^{1-\alpha}}$,
${\mathcal{V}}''=6{{({\kappa_5^2}/3)}^{1-\alpha}}\eta Z\geq 0$, this is controlled by the
sign of $\mathcal{V}$ at its minimum
${{\mathcal{V}}_m}={\mathcal{V}}({Z_m})$ where
${Z_m}=\sqrt{
{{(3k/{\kappa_5^2})}^{1-\alpha}}/3\eta}$. If ${{\mathcal{V}}_m}>0$ then
there are no regular rebounce points and the
collapsing shells may fall from infinity to the singularity at
$S=0$ where $V(0)={\kappa_5^2}{{a}^{1/(1-\alpha)}}/3>0$. For
${{\mathcal{V}}_m}=0$ we have just one regular fixed point $S={S_*}$ which
divides the phase space into two disconnected regions, a bounded
region with the singularity at $S=0$, $0\leq S<{S_*}$, and an
infinitely extended region, $S>{S_*}$,
where the shells may expand increasingly faster to infinity. The
solutions restricted to this region are regular. If
${{\mathcal{V}}_m}<0$ then we have two regular rebounce epochs $S={S_-}$
and $S={S_+}$ such that ${S_-}<{S_+}$. The region
between them is forbidden as there the potential is negative.
For $0\leq S\leq{S_-}$ a shell may expand
to a maximum radius $r{S_-}$ and then rebounce to collapse towards the
singularity at $S=0$. For $S\geq {S_+}$ the collapsing shells shrink
to the minimum scale $S_+$ and then rebounce to expand with ever
increasing speed to infinity. For $\eta<0$ and
$a<0$ we find the same type of dynamics but now only
for the special values $\alpha=-p/q$, $q>p$ with $q$ and $p$
, respectively, even and odd integers.

If $k<0$ then for $\eta>0$ and
$a>0$ the potential is always positive and so there are only singular
solutions without rebouncing points. For $\eta<0$ and $a<0$ we must
consider $\alpha=-p/q$, $q>p$
with $q$ and $p$, respectively, odd and even integers to find
solutions with rebouncing points. The condition $V\geq 0$ is still
equivalent to ${\mathcal{V}}\geq 0$ but now

\beq
{\mathcal{V}}=-{{\left({{\kappa_5^2}\over{3}}\right)}^{1-\alpha}}
\left(|\eta|{Z^3}+|a|\right)+{{|k|}^{1-\alpha}}Z.
\eeq
Now since ${\mathcal{V}}(0)=-{{({\kappa_5^2}/3)}^{1-\alpha}}|a|<0$,
${\mathcal{V}}'=-3{{({\kappa_5^2}/3)}^{1-\alpha}}|\eta|{Z^2}
+{{|k|}^{1-\alpha}}$ and 
${\mathcal{V}}''=-6{{({\kappa_5^2}/3)}^{1-\alpha}}|\eta| Z\leq 0$, the sign of
$\mathcal{V}$ at its maximum
${{\mathcal{V}}_M}={\mathcal{V}}({Z_M})$ where
${Z_M}=\sqrt{
{{(3|k|/{\kappa_5^2})}^{1-\alpha}}/3|\eta|}$
shows that the only
possibilities are the existence of one or two regular
rebounce epochs. In the former case
the classical brane stays forever in the fixed point and in the latter it
oscilates back and forth between the two regular rebounce
epochs.

\section{Conclusions}

In this paper we have investigated if the localized gravitational
dynamics of a polytropic gas could be generated in a RS brane world.
To determine such solutions we have considered matter fields distributed in
the AdS bulk and analyzed the 5-dimensional Einstein
equations using a global conformal transformation whose factor is the
$Z_2$ symmetric warp. Using bulk matter defined by a stress-energy
tensor whose conformal weight is -4 we have found
5-dimensional solutions for which gravity is localized near the brane
and the dynamics of the conformal bulk matter on the
brane is that of an homogeneous polytropic gas. We have analyzed the
polytropic dynamics from the point of view of a 4-dimensional observer
confined to the brane to find conditions for singular or globally
regular behavior. For the examples analyzed we have found that the
dynamical phase space
displaying the evolution of polytropic shells on the brane has an overall
pattern similar to that of homogeneous dark radiation
\cite{RC1}. As we have noted, since the conformal weight is -4 the
bulk matter density and
pressures increase with the coordinate of the fifth dimension. This is
not a problem in a two brane model, but in a single brane model they
diverge at the AdS horizon. Within the single brane model a solution to this
problem requires the simultaneous localization of gravity and matter
near the brane. An analysis of scenarios in which both matter and
gravity are localized on a single brane will be published elsewhere.
\vspace{0.25cm}

\centerline{\bf Acknowledgements}

We are grateful for financial support from the
{\it Funda\c {c}\~ao para a Ci\^encia e a Tecnologia} (FCT) as well
as the {\it Fundo Social Europeu} (FSE) under the contracts
SFRH/BPD/7182/2001 and POCTI/32694/FIS/2000
({\it III Quadro Comunit\'ario de Apoio}).


\begin{thebibliography}{30}

\bibitem{RS1}
L. Randall and R. Sundrum, Phys. Rev. Lett. {\bf 83}, 3370 (1999)
\href{http://arXiv.org/abs/hep-ph/9905221}{[arXiv:hep-ph/9905221]}.

\bibitem{RS2}
L. Randall and R. Sundrum, Phys. Rev. Lett. {\bf 83}, 4690 (1999)
\href{http://arXiv.org/abs/hep-th/9906064}{[arXiv:hep-th/9906064]}.

\bibitem{GW}
W. D. Goldberger and M. B. Wise, Phys. Rev. D. {\bf 60}, 107505 (1999)
\href{http://arXiv.org/abs/hep-ph/9907218}{[arXiv:hep-ph/9907218]};
Phys. Rev. Lett. {\bf 83}, 4922 (1999)
\href{http://arXiv.org/abs/hep-ph/9907447}{[arXiv:hep-ph/9907447]};
Phys. Lett. B {\bf 475}, 275 (2000)
\href{http://arXiv.org/abs/hep-ph/9911457}{[arXiv:hep-ph/9911457]}.

\bibitem{WFGK}
O. DeWolf, D. Z. Freedman, S. S. Gubser and A. Karch, Phys. Rev. D
{\bf 62}, 046008 (2000)
\href{http://arXiv.org/abs/hep-th/9909134}{[arXiv:hep-th/9909134]}.

\bibitem{CGRT}
C. Cs\'aki, M. Graesser, L. Randall and J. Terning, Phys. Rev. D {\bf
  62}, 045015 (2000)
\href{http://arXiv.org/abs/hep-ph/9911406}{[arXiv:hep-ph/9911406]}.

\bibitem{TM}
T. Tanaka and X. Montes, Nucl. Phys. {\bf B582}, 259 (2000)
\href{http://arXiv.org/abs/hep-th/0001092}{[arXiv:hep-th/0001092]}.

\bibitem{LR}
J. Lykken and L. Randall, J. High Energy Phys. {\bf 0006}, 014 (2000)
\href{http://arXiv.org/abs/hep-th/9908076}{[arXiv:hep-th/9908076]}.

\bibitem{GT}
J. Garriga and T. Tanaka, Phys. Rev. Lett. {\bf 84}, 2778 (2000)
\href{http://arXiv.org/abs/hep-th/9911055}{[arXiv:hep-th/9911055]}.

\bibitem{GKR}
S. Giddings, E. Katz and L. Randall, J. High Energy Phys. {\bf 03}, 023 (2000)
\href{http://arXiv.org/abs/hep-th/0002091}{[arXiv:hep-th/0002091]}.

\bibitem{NK}
N. Kaloper, Phys. Rev. D {\bf 60}, 123506 (1999)
\href{http://arXiv.org/abs/hep-th/9905210}{[arXiv:hep-th/9905210]};

T. Nihei, Phys. Lett. B {\bf 465}, 81 (1999)
\href{http://arXiv.org/abs/hep-ph/9905487}{[arXiv:hep-ph/9905487]};

C. Cs\'aki, M. Graesser, C. Kolda and J. Terning, Phys. Lett. B {\bf
462}, 34 (1999)
\href{http://arXiv.org/abs/hep-ph/9906513}{[arXiv:hep-ph/9906513]};

J. M. Cline, C. Grojean and G. Servant, Phys. Rev. Lett. {\bf 83},
4245 (1999)
\href{http://arXiv.org/abs/hep-ph/9906523}{[arXiv:hep-ph/9906523]}.

\bibitem{KKOP}
P. Kanti, I. I. Kogan, K. A. Olive and M. Pospelov, Phys. Lett. B {\bf
  468}, 31 (1999)
\href{http://arXiv.org/abs/hep-ph/9909481}{[arXiv:hep-ph/9909481]};
Phys. Rev. D {\bf 61}, 106004 (2000)
\href{http://arXiv.org/abs/hep-ph/9912266}{[arXiv:hep-ph/9912266]};

\bibitem{CHR}
A. Chamblin, S. W. Hawking and H. S. Reall, Phys. Rev. D {\bf 61}, 065007
(2000)
\href{http://arXiv.org/abs/hep-th/9909205}{[arXiv:hep-th/9909205]}.

\bibitem{EHM}
R. Emparan, G. T. Horowitz and R. C. Myers, J. High Energy Phys. {\bf
  0001}, 007 (2000)
\href{http://arXiv.org/abs/hep-th/9911043}{[arXiv:hep-th/9911043]};
J. High Energy Phys. {\bf 0001}, 021 (2000)
\href{http://arXiv.org/abs/hep-th/9912135}{[arXiv:hep-th/9912135]}.

\bibitem{KOP}
P. Kanti, K. A. Olive and M. Pospelov, Phys. Lett. B {\bf 481}, 386
(2000)
\href{http://arXiv.org/abs/hep-ph/0002229}{[arXiv:hep-ph/0002229]}.

\bibitem{SS}
T. Shiromizu and M. Shibata, Phys. Rev. D {\bf 62}, 127502 (2000)
\href{http://arXiv.org/abs/hep-th/0007203}{[arXiv:hep-th/0007203]}.

\bibitem{LP}
H. L\"u and C. N. Pope, Nucl. Phys. {\bf B598}, 492 (2001)
\href{http://arXiv.org/abs/hep-th/0008050}{[arXiv:hep-th/0008050]}.

\bibitem{CRSS}
A. Chamblin, H. R. Reall, H. a. Shinkai and T. Shiromizu, Phys. Rev. D
{\bf 63}, 064015 (2001)
\href{http://arXiv.org/abs/hep-th/0008177}{[arXiv:hep-th/0008177]}.

\bibitem{MPS}
M. S. Modgil, S. Panda and G. Sengupta, Mod. Phys. Lett. A {\bf 17}
1479 (2002)
\href{http://arXiv.org/abs/hep-th/0104122}{[arXiv:hep-th/0104122]}.

\bibitem{KT}
P. Kanti and K. Tamvakis, Phys. Rev. D {\bf 65}, 084010 (2002)
\href{http://arXiv.org/abs/hep-th/0110298}{[arXiv:hep-th/0110298]}.

\bibitem{RC}
R. Neves and C. Vaz, {\it Brane World Dynamics and Conformal Bulk
  Fields}, Phys. Rev. D (in press) 
\href{http://arXiv.org/abs/hep-th/0302030}{[arXiv:hep-th/0302030]}.

\bibitem{KMP}
A. Kamenshchik, U. Moschella and V. Pasquier, Phys. Lett. B {\bf 511},
265 (2001)\href{http://arXiv.org/abs/gr-qc/0103004}{[arXiv:gr-qc/0103004]}.

\bibitem{BTV}
N. Bili\'c, G. B. Tupper and R. D. Viollier, Phys. Lett. B {\bf 535},
17 (2002)
\href{http://arXiv.org/abs/astro-ph/0111325}{[arXiv:astro-ph/0111325]}.

\bibitem{BBS}
M. C. Bento, O. Bertolami and S. S. Sen, Phys. Rev. D {\bf 66}, 043507 (2002)
\href{http://arXiv.org/abs/gr-qc/0202064}{[arXiv:gr-qc/0202064]};
Phys. Rev. D {\bf 67}, 063003 (2003)
\href{http://arXiv.org/abs/astro-ph/0210468}{[arXiv:astro-ph/0210468]}.

\bibitem{IR}
I. Z. Rothstein, Phys. Rev. D {\bf 64}, 084024 (2001)
\href{http://arXiv.org/abs/hep-th/0106022}{[arXiv:hep-th/0106022]}.

\bibitem{SMS}
T. Shiromizu, K. I. Maeda and M. Sasaki, Phys. Rev. D {\bf 62}, 024012
(2000)
\href{http://arXiv.org/abs/gr-qc/9910076}{[arXiv:gr-qc/9910076]}.

\bibitem{RM}
R. Maartens, Phys. Rev. D {\bf 62}, 084023 (2000)
\href{http://arXiv.org/abs/hep-th/0004166}{[arXiv:hep-th/0004166]}.

\bibitem{RC1}
R. Neves and C. Vaz, Phys. Rev. D {\bf 66}, 124002 (2002)
\href{http://arXiv.org/abs/hep-th/0207173}{[arXiv:hep-th/0207173]}.

\bibitem{DJCJ}
S. S. Deshingkar, S. Jhingan, A. Chamorro and P. S. Joshi, Phys.
Rev. D {\bf 63}, 124005 (2001)
\href{http://arXiv.org/abs/gr-qc/0010027}{[arXiv:gr-qc/0010027]}.
\end{thebibliography}
\end{document}